\documentclass[a4paper,11pt]{article}
\usepackage{pos}
\usepackage{enumitem}
\usepackage{multirow}
\newcommand{\tr}{\operatorname{Tr}}

\NewDocumentCommand\Nf{mgg}{$N_{\textrm{f}}\text{=}#1\IfNoValueTF{#2}{}{\text{+}#2}\IfNoValueTF{#3}{}{\text{+}#3}$}

\title{Nucleon form factors from \Nf{2}{1}{1} twisted mass QCD at the physical point}
\ShortTitle{Nucleon FFs from \Nf{2}{1}{1} tmQCD}

\author[a,b]{C.~Alexandrou}
\author[b]{S.~Bacchio}
\author[c]{M.~Constantinou}
\author[b]{J.~Finkenrath}
\author[a,b]{K.~Hadjiyiannakou}
\author[d]{K.~Jansen}
\author*[b]{G.~Koutsou}
\author[e]{A.~Vaquero}

\affiliation[a]{Department of Physics, University of Cyprus,
  P.O. Box 20537, 1678 Nicosia, Cyprus}

\affiliation[b]{Computation-based Science and Technology Research Center, The Cyprus Institute,
  20 Konstantinou Kavafi Str., 2121, Nicosia Cyprus}

\affiliation[c]{Temple University, 1801 N Broad Str., Philadelphia, PA 19122, USA}

\affiliation[d]{DESY-Zeuthen, 6 Platanenallee Str., 15738 Zeuthen, Germany}

\affiliation[d]{University of Utah, 201 Presidents' Cir, Salt Lake City, UT 84112, USA}

\emailAdd{g.koutsou@cyi.ac.cy}

\abstract{We present the nucleon axial and electromagnetic form
  factors using \Nf{2}{1}{1} ensembles of twisted mass fermions with
  clover improvement and with masses tuned to their physical
  values. Excited state effects are studied using several sink-source
  time separations in the range 0.8 fm - 1.6 fm, exponentially
  increasing statistics with the separation such that statistical
  errors remain approximately constant. In addition, quark loop
  disconnected diagrams are included in order to extract the isoscalar
  axial form factors and the proton and neutron electromagnetic form
  factors, as well as their strange-quark contributions. The radii and
  moments are extracted by modelling the $Q^2$ dependence, including
  using the so-called $z$-expansion. A preliminary assessment of
  lattice cut-off effects is presented using two lattice spacings
  directly at the physical point.  }

\FullConference{%
 The 38th International Symposium on Lattice Field Theory, LATTICE2021
  26th-30th July, 2021
  Zoom/Gather@Massachusetts Institute of Technology
}

\begin{document}
\maketitle

\section{Introduction}
Nucleon form factors are fundamental probes of its structure, mapping
for instance the charge distribution of its constituent quarks. In
particular, the electromagnetic form factors determine the nucleon
magnetic moment and the electric and magnetic radii, while the axial
form factors probe chiral symmetry and test partial conservation of
the axial current (PCAC). Electron scattering experiments can provide
a precise determination of the nucleon electromagnetic form factors.
Experiments, having been carried out since the fifties, are continuing
at experimental facilities at Mainz and JLab. In the limit of zero
momentum transfer $Q^2$, the slope of the electric, $G_E(Q^2)$, and
magnetic, $G_M(Q^2)$, form factors is related to the electric and
magnetic root mean square (rms) radii.  Their value at $Q^2=0$ yields
the electric charge and magnetic moment, respectively. Furthermore,
the process $\nu_\mu + p \rightarrow \mu^+ +n$ yields the axial form
factor, $G_A(Q^2)$, while the induced pseudoscalar form factor,
$G_P(Q^2)$, is obtained via the longitudinal cross section in pion
electro-production. The nucleon axial charge, $g_A=G_A(0)$ is measured
to high precision from $\beta$-decay experiments.

In this contribution, we present a calculation of the electromagnetic
and axial form factors of the nucleon using lattice QCD on three
ensembles of twisted mass clover-improved fermion ensembles with quark
masses tuned to yield physical pion mass values (physical point). Two
ensembles are simulated with two degenerate light quarks (\Nf{2}),
lattice spacing $a$=0.0937~fm and three-dimensional volumes
$L^3\simeq$(4.5~fm)$^3$ and $L^3\simeq$(6~fm)$^3$, while the third
ensemble is simulated with a strange and charm quark in addition
(\Nf{2}{1}{1}), $a$=0.0801~fm and $L^3\simeq$(5.1~fm)$^3$. The
calculation of the disconnected quark loop contributions allows the
extraction of the individual proton and neutron form factors, as well
as the strange form factors. Furthermore, preliminary results are
presented for the strange electromagnetic form factors on a physical
point \Nf{2}{1}{1} ensemble with $a\simeq$0.07~fm and
$L^3\simeq$(5.6~fm)$^3$.

\section{Lattice setup}
\subsection{Matrix elements}
The form factors are obtained from the nucleon matrix elements:
\[
\langle N(p', s')|\mathcal{O}^{X}_\mu|N(p,s)\rangle =
\sqrt{\frac{m^2_N}{E_N(\vec{p}')E_N(\vec{p})}}\bar{u}_N(p',s')\Lambda^{X}_\mu(q^2)u_N(p,s)
\]
with $N(p,s)$ a nucleon state of momentum $p$ and spin $s$,
$E_N(\vec{p}) = p_0$ its energy and $m_N$ its mass, $u_N$ a nucleon
spinor and, $q=p'-p$, the momentum transfer from initial ($p$) to
final ($p'$) momentum, and $\mathcal{O}^{X}$ either vector ($X=V$) or
axial ($X=A$) current. The nucleon matrix element of the vector
current decomposes into the Dirac $F_1$ and Pauli $F_2$ form factors,
while the corresponding one of the axial current decomposes into the
axial $G_A$ and induced pseudo-scalar $G_p$ form factors, given as
\begin{align}
  \Lambda_\mu^V(q^2) = \gamma_\mu
  F_1(q^2)+\frac{i\sigma_{\mu\nu}q^\nu}{2m_N}F_2(q^2),\quad
  \Lambda^A_\mu(q^2) = \frac{i}{2}\gamma_5\gamma_\mu
  G_A(q^2)+\frac{q_\mu\gamma_5}{2m_N}G_p(q^2).
\end{align}
The Dirac and Pauli form factors can also be expressed in terms of the
nucleon electric $G_E$ and magnetic $G_M$ Sachs form factors via
$G_E(q^2) = F_1(q^2)+\frac{q^2}{(2m_N)^2}F_2(q^2)$ and $G_M(q^2) =
F_1(q^2)+F_2(q^2)$.

\subsection{Lattice extraction of form factors}
\label{sec:extraction}
On the lattice, the required matrix elements are obtained from
combinations of two- and three-point correlation functions,
\begin{align}
  C(\Gamma_0,\vec{p};t_s,t_0) {=} &\sum_{\vec{x}_s} \hspace{-0.1cm} \tr
  \left[ \Gamma_0 {\langle}J_N(x_s) \bar{J}_N(x_0) {\rangle}
    \right]e^{{-}i (\vec{x}_s{-}\vec{x}_0) \cdot \vec{p}}\quad\textrm{and}
  \label{eq:two-point}\\
C_\mu(\Gamma_\nu,\vec{p},\vec{p}';t_s,t_{\rm ins},t_0) {=}&
{\sum_{\vec{x}_{\rm ins},\vec{x}_s}} e^{i (\vec{x}_{\rm ins} {-}
  \vec{x}_0) \cdot \vec{q}} e^{-i(\vec{x}_s {-} \vec{x}_0)\cdot
  \vec{p}\,'} \tr \left[ \Gamma_\nu \langle J_N(x_s) j_\mu(x_{\rm
    ins}) \bar{J}_N(x_0) \rangle \right]
  \label{eq:three-point}
\end{align}
respectively, with $J$ the interpolating field of the nucleon,
$x_0=(t_0, \vec{x}_0)$ the \textit{source}, $j_\mu$ the
electromagnetic or axial current, $x_{\rm ins}=(t_{\rm ins},
\vec{x}_{\rm ins})$ the \textit{insertion}, and $x_s=(t_s, \vec{x}_s)$
the \textit{sink}. $\Gamma_\nu$ is a projector acting on spin indices,
with $\Gamma_0 {=} \frac{1}{2}(1{+}\gamma_0)$ and $\Gamma_k{=}\Gamma_0
i \gamma_5 \gamma_k$.

We form a ratio of three- to two-point
functions~\cite{Alexandrou:2010hf} so as to cancel unknown overlaps
and energy exponentials and, after taking the large time limit, yields
the nucleon ground state matrix element,
$\Pi_\mu(\Gamma_\nu;\vec{p},\vec{p}')$, namely
$R_\mu(\Gamma_\nu;\vec{p},\vec{p}';t_s;t_{\rm ins})\xrightarrow[t_{\rm
    ins}\gg]{t_s-t_{\rm ins}\gg}\Pi_\mu(\Gamma_\nu;\vec{p},\vec{p}')$.
Since statistical errors increase exponentially with time separation,
we cannot take $t_\mathrm{ins}$ and $t_s$ arbitrarily large and,
therefore, we evaluate the convergence to the ground state using the following methods:
\begin{itemize}[leftmargin=*]
  \setlength\itemsep{-0.25em}
  \item \emph{Plateau method:} We identify a time-independent window
    (plateau) as a function of $t_{\rm ins}$ and fit to extract the plate value. We seek convergence  of the plateau value as we  increase $t_s$ to extract  the desired matrix element.

  \item \emph{Two-state  fit method:} We fit the two- and three-point
    functions considering contributions up to the first excited state,
    i.e.  using the expressions
    \begin{align}
      C(\vec{p},t_s) =& \sum_{i=0}^1c_i(\vec{p}) e^{-E_i(\vec{p}) t_s}\quad\textrm{and}\label{eq:twop two-state fit}\\
      C_\mu(\Gamma_\nu,\vec{p},\vec{p}',t_s,t_{\rm ins}) =&
      \sum_{i,j=0}^{1}A_{ij}^\mu(\Gamma_\nu,\vec{p},\vec{p}')
      e^{-E_i(\vec{p}\,')(t_s-t_{\rm ins})-E_j(\vec{p})t_{\rm ins}},\label{eq:thrp three-state-fit}
    \end{align}
    where $A^\mu_{ij}$ are proportional to the matrix element $\langle
    i | O_\mu | j \rangle$, with $|0\rangle$ and $|1\rangle$ denoting
    the ground and first exited state and $E_0$ and $E_1$ their
    energies, respectively. The desired matrix element is obtained via
    $\Pi_\mu(\Gamma_\nu;\vec{p},\vec{p}')=\frac{A_{00}^\mu(\Gamma_\nu,\vec{p}\,',\vec{p})}{\sqrt{c_0(\vec{p}\,')
        c_0(\vec{p})}}$.
  \item \emph{Summation method:} we sum the ratio over $t_{\rm ins}$,
    \cite{Maiani:1987by,Capitani:2012gj} which for large $t_s$ yields:
    $R^{\rm sum}_\mu(\Gamma_\nu;\vec{p},\vec{p}';t_s)=\sum_{t_{\rm
        ins}}R_\mu(\Gamma_\nu;\vec{p},\vec{p}';t_s;t_{\rm
      ins})\xrightarrow{t_s\gg}c + t_s
    \Pi_\mu(\Gamma_\nu;\vec{p},\vec{p}')$. We carry out a linear fit
    with $t_s$ in order to extract the desired matrix element. We will
    also use the \emph{derivative summation method}, obtained by
    taking the finite difference of $R^{\rm sum}_\mu$ and fitting to a
    constant.
\end{itemize}

For the connected three-point functions, we use sequential inversions
through the sink fixing the sink momentum $\vec{p}'$ to zero, which
constrains $\vec{p}=-\vec{q}$. Having $\Pi_\mu(\Gamma;\vec{q})$,
different combinations of current insertion directions ($\mu$) and
projections  $\Gamma_\mu$ yield different form
factors. Using $\Pi^V$ to denote electromagnetic and $\Pi^A$ for axial
matrix elements, we have:
\begin{align}
  \Pi^V_0(\Gamma_0;\vec{q}) = & \mathcal{C}\frac{E_N+m_N}{2m_N}G_E(Q^2),\quad          & \Pi^A_i(\Gamma_k;\vec{q}) = & \frac{i\mathcal{C}}{4m_N}[\frac{q_kq_i}{2m_N}G_p(Q^2) - (E_N+m_N)\delta_{ik}G_A(Q^2)],\nonumber \\
  \Pi^V_i(\Gamma_k;\vec{q}) = & \mathcal{C}\frac{\epsilon_{ijk}q_j}{2m_N}G_M(Q^2),\quad & \Pi^A_0(\Gamma_k;\vec{q}) = & C\frac{-q_k}{2m_N}[G_A(Q^2) + G_P(Q^2)\frac{m_N-E_N}{2m_N}], \textrm{and} \nonumber\\
  \Pi^V_i(\Gamma_0;\vec{q}) = & \mathcal{C}\frac{q_i}{2m_N}G_E(Q^2),\label{eq:ffs} 
\end{align}
where $Q^2=-q^2$, $\mathcal{C}=\sqrt{\frac{2m_N^2}{E_N(E_N+m_N)}}$,
and the projectors $\Gamma_0=\frac{1+\gamma_0}{4}$ and
$\Gamma_k=i\gamma_5\gamma_k\Gamma_0$, with $i,k=1,2,3$. For the
disconnected contributions, finite values of the sink momentum
$\vec{p}'\neq 0$ are used to increase statistics, with details of the
setup provided in Refs.~\cite{Alexandrou:2019olr, Alexandrou:2021wzv}.

\subsection{Lattice setup}
We use two \Nf{2} ensembles simulated using twisted mass
clover-improved fermions and two lattice volumes as indicated in
Table~\ref{tab:stat}. Simulation details for these ensembles can be
found in Refs.~\cite{ETM:2015ned,Alexandrou:2018egz}. Two more recent
ensembles at the physical point, using \Nf{2}{1}{1} twisted mass
clover-improved fermions have also been analyzed and are shown in
Table~\ref{tab:stat}, with details on their simulation provided in
Ref.~\cite{Alexandrou:2018egz}.

\begin{table}[h]  
  \begin{minipage}{0.55\linewidth}
    \caption{Bottom: details of the \Nf{2}~\cite{ETM:2015ned} and
      \Nf{2}{1}{1}~\cite{Alexandrou:2018egz} ensembles used. Right:
      Statistics used for each sink-source separation for the case of
      the \texttt{cB211.64} ensemble.}\label{tab:stat} \centering
      \begin{tabular}[b]{cr@{\texttt{.}}lr@{$\times$}lr}\hline\hline
        $N_{\rm f}$               & \multicolumn{2}{c}{Ens. ID} & \multicolumn{2}{c}{Vol.} & $a$ [fm]             \\\hline
        2                         & \texttt{cA2}                & \texttt{48}              & 48$^3$ & 96  & 0.0937 \\                 
        2                         & \texttt{cA2}                & \texttt{64}              & 64$^3$     & 128 & 0.0937 \\                 
        2+1+1                     & \texttt{cB211}              & \texttt{64}              & 64$^3$     & 128 & 0.0801 \\                 
        2+1+1                     & \texttt{cC211}              & \texttt{80}              & 80$^3$     & 160 & 0.070 \\\hline\hline            
      \end{tabular}
  \end{minipage}
  \hfill
  \begin{minipage}{0.4\linewidth}
    \begin{tabular}[b]{rr@{$\times$}l}\hline\hline
      \multirow{2}{*}{$t_s/a$}    & \multicolumn{2}{c}{$N_\textrm{src}\times N_\textrm{conf}$}                    \\
                                  & \multicolumn{2}{c}{\texttt{cB211.64}}                                         \\ \hline                       
      12                          & 4                           & 750                                             \\
      14                          & 6                           & 750                                             \\
      16                          & 16                          & 750                                             \\
      18                          & 48                          & 750                                             \\
      20                          & 64                          & 750                                             \\
      \multicolumn{1}{l}{2-point} & 264                         & 750                                             \\\hline\hline
    \end{tabular}    
  \end{minipage}
\end{table}

\begin{figure}[h]
  \centering\includegraphics[width=0.8\linewidth]{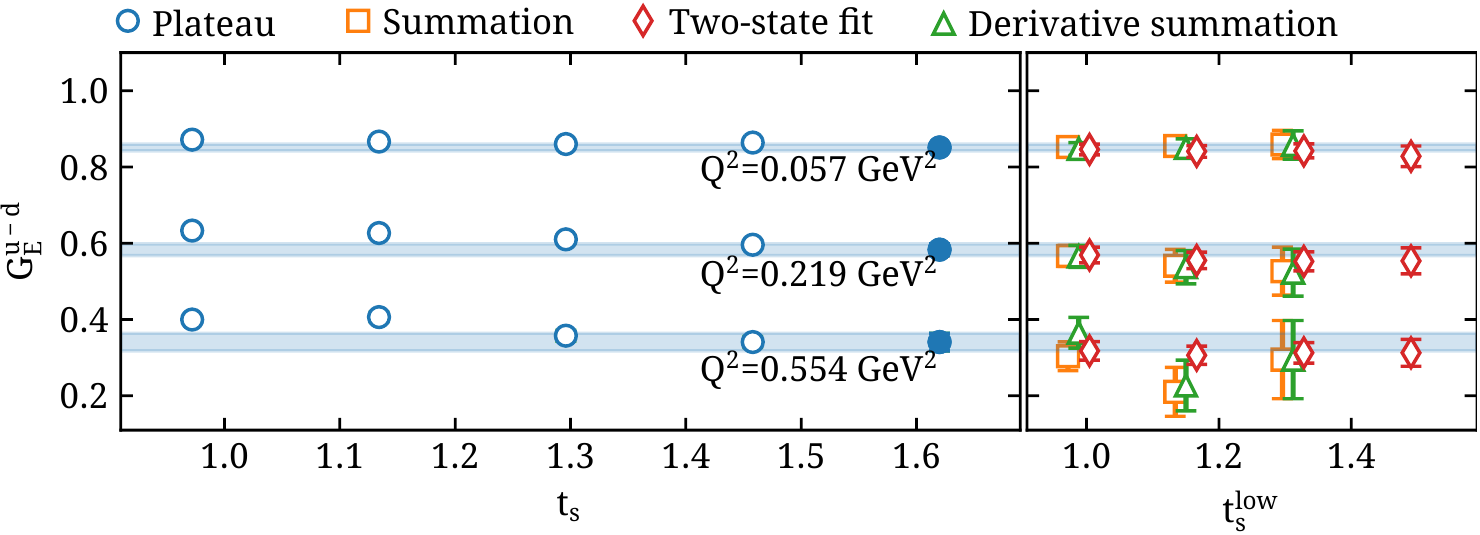}
  \caption{Plateau (circles), summation (squares), and two-state fit
    (rhombuses) methods used for the isovector contribution to
    $G_E(Q^2)$ for the \texttt{cB211.64} ensemble for three values of
    $Q^2$ as indicated in the plot. $t_s^\textrm{low}$ indicates the
    smallest sink-source separation used in the fit of either
    method. The filled symbol and band spanning the horizontal extent
    of the plots is chosen as the quoted value.}\label{fig:fits}
\end{figure}

For the connected three-point functions, for which we use the fixed
sink sequential inversion approach, we increase statistics with
increasing sink-source separations, as shown in Table~\ref{tab:stat}
for the case of the \texttt{cB211.64} ensemble. The multiple
separations allow an analysis of excited states with the methods of
Sec.~\ref{sec:extraction} as demonstrated for an indicative example in
Fig.~\ref{fig:fits}.

For the disconnected quark loops, we use a combination of eigenvalue
deflation~\cite{Gambhir:2016uwp}, hierarchical
probing~\cite{Stathopoulos:2013aci}, and spin and color
dilution~\cite{McNeile:2006bz}, as explained in
Refs.~\cite{Alexandrou:2018sjm, Alexandrou:2019olr,
  Alexandrou:2021wzv}. For the connected electromagnetic form factors,
we use the conserved vector current, which does not require
renormalization, while for the disconnected case, we use the local
vector current. The renormalization for the local vector current is
carried out non-perturbatively in the RI'-MOM
scheme~\cite{Martinelli:1994ty} employing momentum sources, following
the procedures described in
Refs.~\cite{Alexandrou:2010me,Alexandrou:2015sea}. For the axial
current, both singlet and non-singlet, are computed. For the case of
the disconnected contributions, we increase statistics using
additional two-point functions as indicated in the last row of the
right panel of Table~\ref{tab:stat}. Details regarding the methods
used for obtaining the disconnected contributions can be found in
Refs.~\cite{Alexandrou:2018sjm} and~\cite{Alexandrou:2019olr}. For the
\Nf{2} ensembles, details for the separations and statistics used can
be found in Refs.~\cite{Alexandrou:2017ypw,Alexandrou:2018lvq}.

\section{Results}
\subsection{Electromagnetic form factors}
\begin{figure}[h]
  \includegraphics[width=\linewidth]{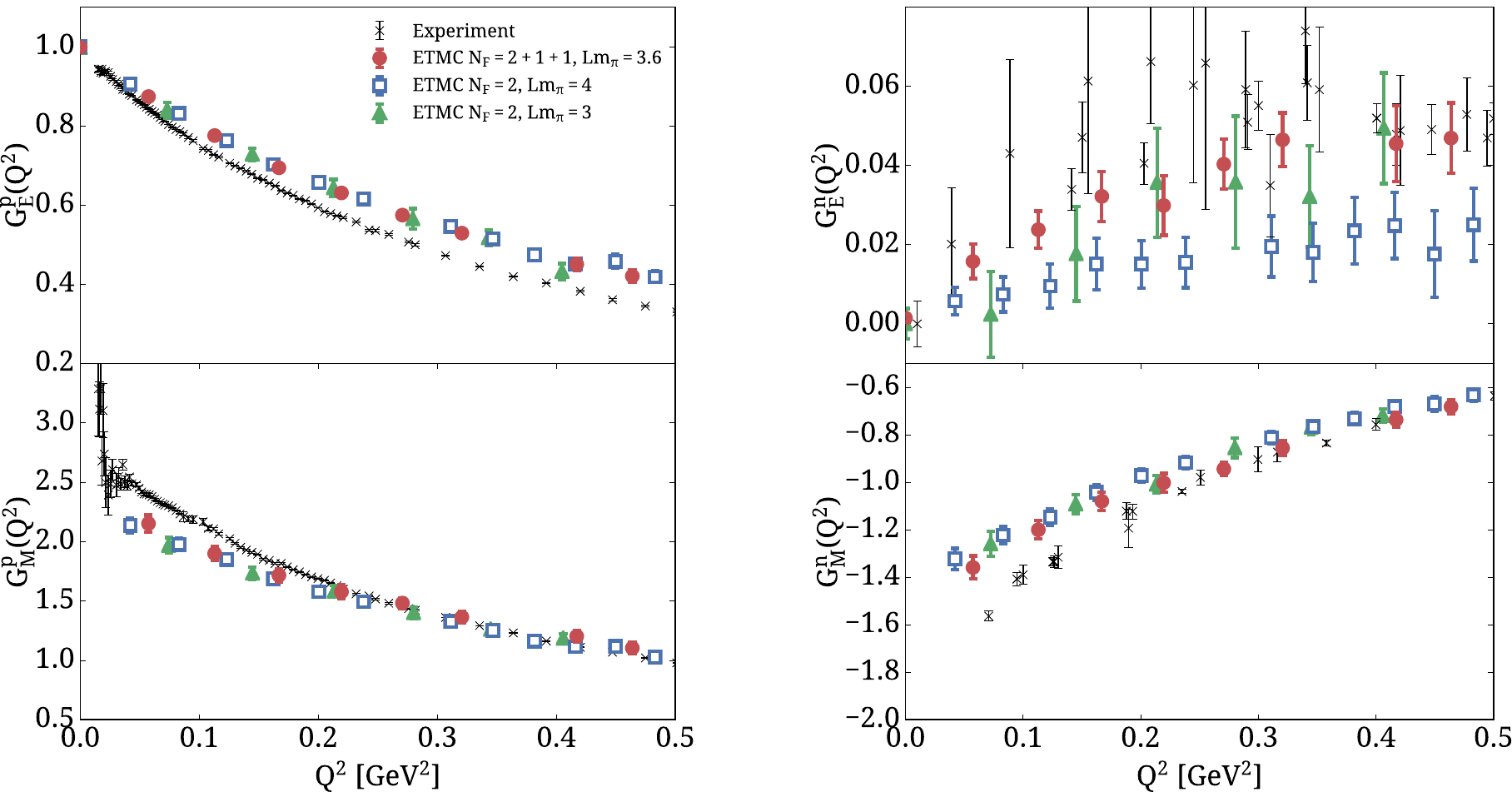}
  \caption{The proton (left) and neutron (right) electric (top) and
    magnetic (bottom) form factors for the two \Nf{2} ensembles of
    Table~\ref{tab:stat} (green triangles and blue square) and for the
    \texttt{cB211.64} ensemble (red circles). Filled symbols indicate
    that disconnected contributions have been
    included.}\label{fig:gempn}
\end{figure}

\begin{figure}
  \includegraphics[width=\linewidth]{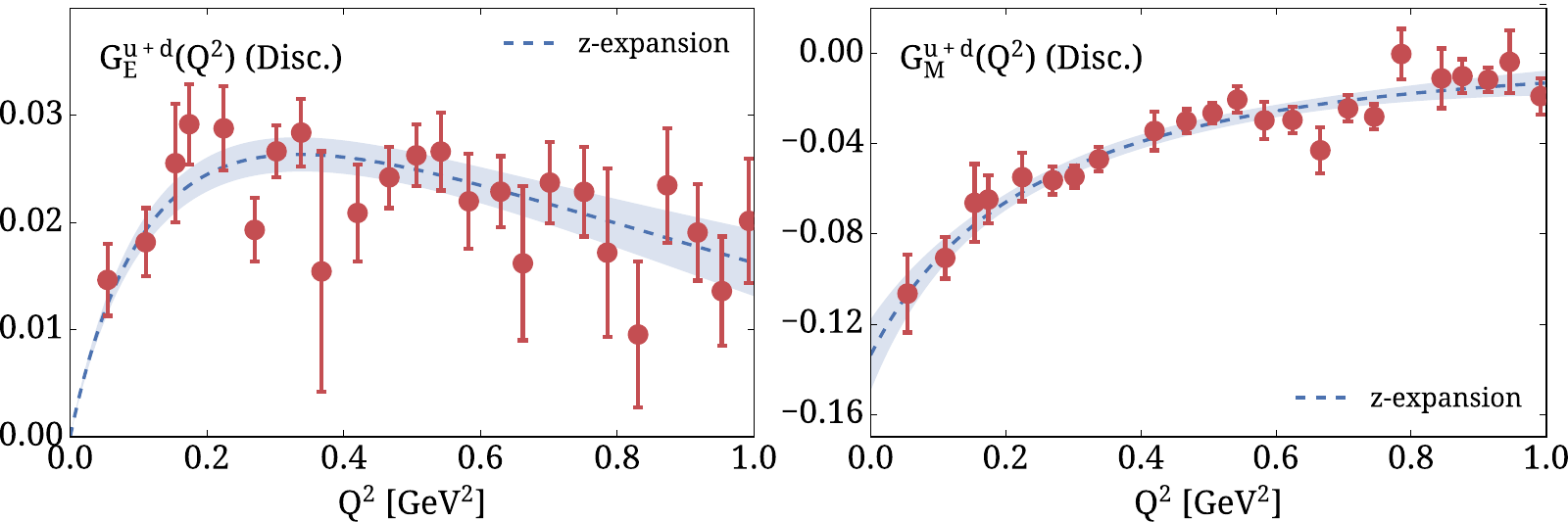}
  \caption{Disconnected contributions to the isoscalar electric (left)
    and magnetic (right) form factors of the nucleon for the
    \texttt{cB211.64} ensemble. The band is a fit to the
    $z$-expansion.}\label{fig:gemdisc}
\end{figure}

The proton and neutron Sachs electric ($G_E^p(Q^2)$ and $G_E^n(Q^2)$)
and magnetic ($G_M^p(Q^2)$ and $G_M^n(Q^2)$) form factors are obtained
in the flavor isospin limit by combining the isovector
($G_{E/M}^{v}(Q^2)$) and isoscalar ($G_{E/M}^{s}(Q^2)$) cases, via
$G_{E/M}^p(Q^2) = \frac{1}{2}[G_{E/M}^{v}(Q^2) +
  \frac{1}{3}G_{E/M}^{s}(Q^2)]$ and $G_{E/M}^n(Q^2) =
\frac{1}{2}[\frac{1}{3}G_{E/M}^{s}(Q^2) -
  G_{E/M}^{v}(Q^2)]$. Disconnected quark loop contributions cancel for
the isovector case but need to be computed for the isoscalar case. In
Fig.~\ref{fig:gempn} we show results for the proton and neutron Sachs
electromagnetic form factors for the two \Nf{2} ensembles and the
\texttt{cB211.64} \Nf{2}{1}{1} ensemble. Disconnected contributions
are included in the results for the latter and
the \texttt{cA2.48} ensemble. Comparison between the two \Nf{2}
ensembles suggests that volume effects are within statistical errors
for the range of volumes used here. Furthermore, our high accuracy
determination of the disconnected contributions at the physical point
reveals that these are at the percent level compared to connected
contributions, as can be verified from Fig.~\ref{fig:gemdisc} in which
we plot them separately for the \texttt{cB211.64} ensemble. In Fig.~\ref{fig:gemdisc} we also show fits to the so-called
 $z$-expansion~\cite{Hill:2010yb}:
\begin{equation}
    G(Q^2) = \sum_{k=0}^{k_{\rm max}} a_k z^k,\qquad z=\frac{\sqrt{t_{\rm cut}+Q^2} - \sqrt{t_{\rm cut}}}{\sqrt{t_{\rm cut}+Q^2} + \sqrt{t_{\rm cut}}},
\end{equation}
where we use $G(Q^2)$ to refer to either electric or magnetic form
factor and the values for $t_\textrm{cut}$, the choice of
$k_\textrm{max}$ and the priors used for the fit parameters $a_k$ are
as explained in Ref.~\cite{Alexandrou:2018sjm}.

\begin{figure}
  \includegraphics[width=\linewidth]{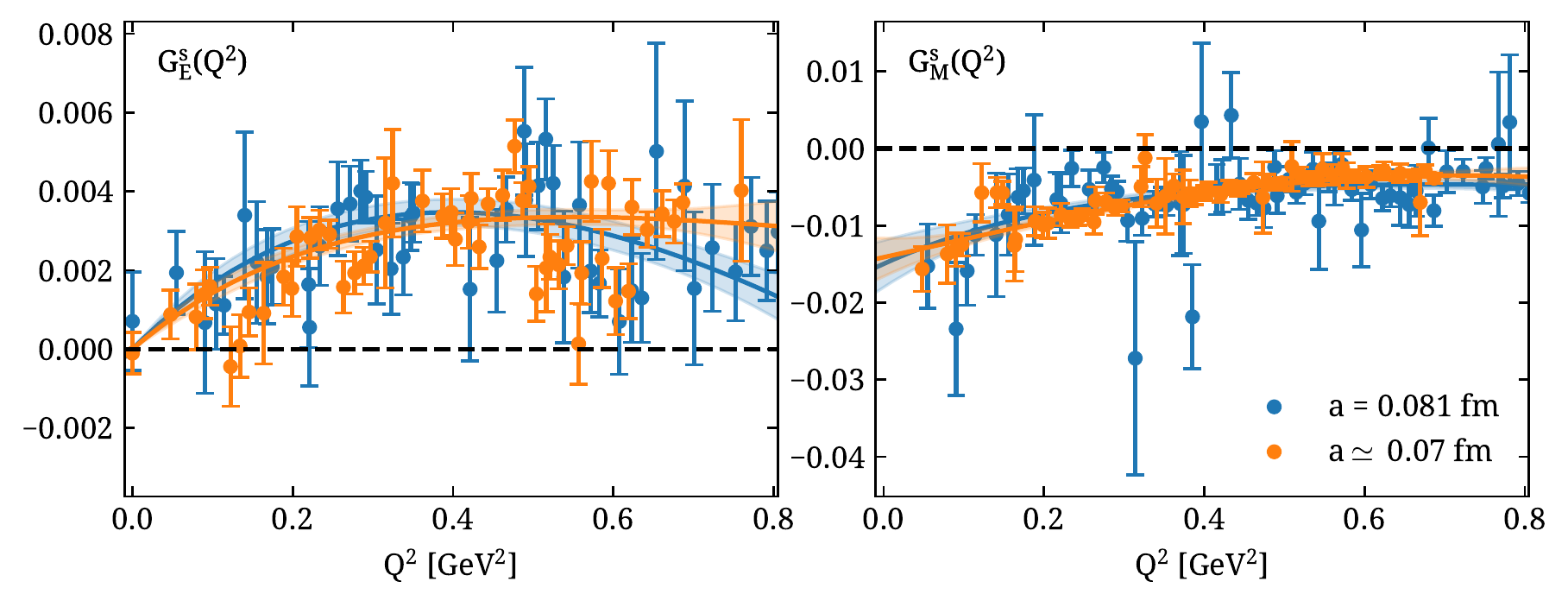}
  \caption{Strange electric (left) and magnetic (right) form factors
    obtained on the \texttt{cB211.64} (blue circles) and
    \texttt{cC211.80} (orange circles) ensembles. The curves are from
    fits to the $z$-expansion. Details on the results obtained for
    \texttt{cB211.64} can be found in Ref.~\cite{Alexandrou:2019olr},
    while results for \texttt{cC211.80} are
    preliminary.}\label{fig:gems}
\end{figure}

The strange electromagnetic form factors come entirely from disconnected
contributions and are computed for the \Nf{2}{1}{1} ensembles to
unprecedented accuracy at the physical point. We show results in
Fig.~\ref{fig:gems}, where the results for the \texttt{cC211.80}
ensemble are preliminary. Indicatively, we note that our results are
compatible recent experimental determinations by
HAPPEX~\cite{HAPPEX:2011xlw} $G^s_M(Q^2{\sim}$0.62~GeV$^2)=$
$-0.070(67)$ and by A4~\cite{Baunack:2009gy}
$G_E^s(Q^2{\sim}$0.22~GeV$^2)=$ $0.050(38)(19)$,
$G_M^s(Q^2{\sim}$0.22~GeV$^2)=$ $0.14(11)(11)$ while our errors are at
the $\sim$25\% level, allowing us to connect to more accurate,
upcoming determinations by Q-weak~\cite{Armstrong:2012ps} experiment
and MESA~\cite{Becker:2018ggl}.

\begin{figure}[h]
  \begin{minipage}[c]{0.5\linewidth}
    \includegraphics[width=\linewidth]{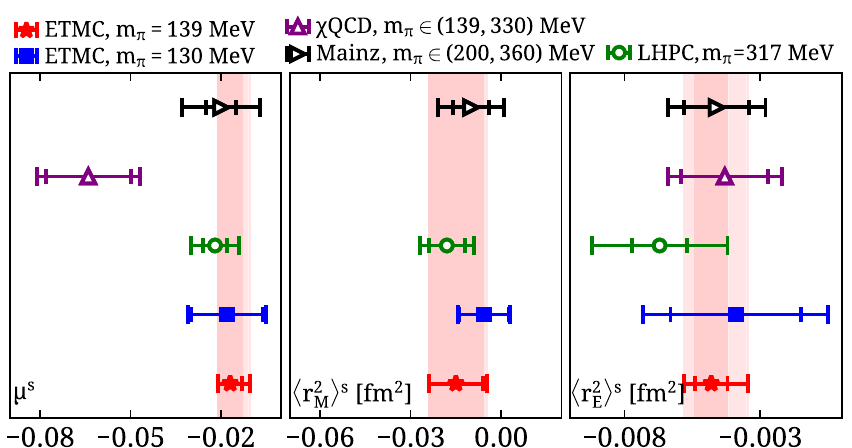}
  \end{minipage}
  \hfill
  \begin{minipage}[c]{0.48\linewidth}
    \caption{The nucleon strange magnetic moment (left) and magnetic
      (center) and electric (right) radii obtained on the
      \texttt{cB211.64} (red stars) and \texttt{cA2.48} (blue circles)
      ensembles. We compare to results from other lattice QCD
      calculations, namely Ref.~\cite{Green:2015wqa} (green circles),
      Ref.~\cite{Sufian:2016pex} (purple triangles), and Ref.~\cite{Djukanovic:2019jtp}
      (black right-pointing triangles).}\label{fig:remus}
  \end{minipage}
\end{figure}

The magnetic form factor at $Q^2{=}0$ yields the magnetic moment
$\mu_s = G_M^s(Q^2{=}0)$, while the slope of the form factors yields
the associated radii, $\langle r_{E/M}^2\rangle^s=-6\frac{\partial
}{\partial Q^2}G^s_{E/M}(Q^2)|_{Q^2=0}$. From fits to the
$z$-expansion, the radii and magnetic moment relate to the fitted
parameters $a_k$ via $\langle r_{E/M}^2\rangle^s =
\frac{-3a^{E/M}_1}{2t_\textrm{cut}}$ and $\mu_s = a^M_0$, where
$a^E_k$ ($a^M_k$) are obtained via $z$-expansion fits to the electric
(magnetic) form factors. In Fig.~\ref{fig:remus} we plot the radii and
magnetic moment for the \texttt{cA2.48} and \texttt{cB211.64}
ensembles and compare to other lattice QCD results. Details on the
fits can be found in Ref.~\cite{Alexandrou:2019olr}.

\subsection{Axial form factors}
\begin{figure}[h]
  \includegraphics[width=0.5\linewidth]{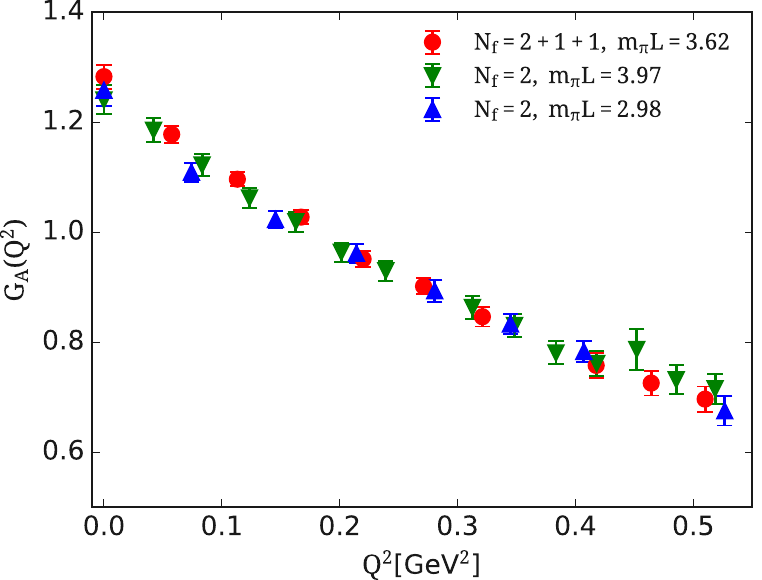}
  \includegraphics[width=0.5\linewidth]{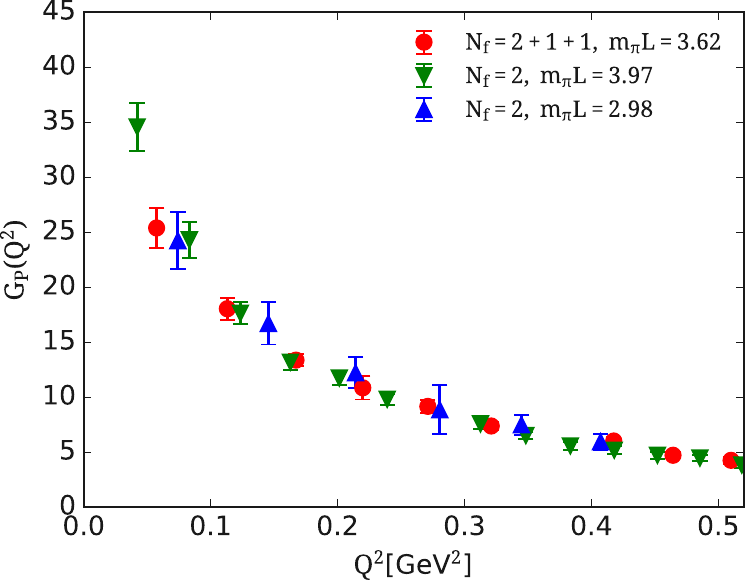}
  \caption{Isovector axial (left) and induced pseudoscalar (right)
    form factors of the nucleon from the \texttt{cB211.64} (red
    circles), \texttt{cA2.48} (blue triangles), and \texttt{cA2.64}
    (green down-pointing triangles) ensembles. }\label{fig:GAv}
\end{figure}

\begin{figure}[h]
  \begin{minipage}{0.4\linewidth}
    \includegraphics[width=\linewidth]{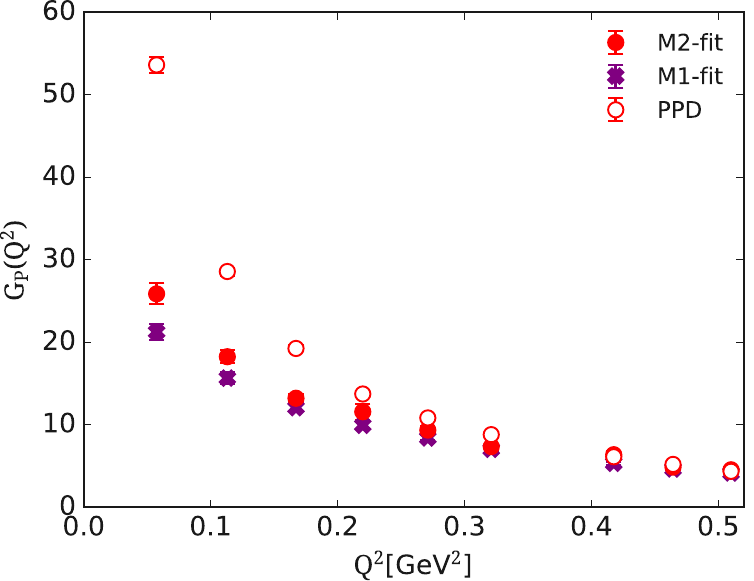}
  \end{minipage}
  \hfill
  \begin{minipage}{0.58\linewidth}
    \caption{The induced pseudoscalar form factor, $G^{u-d}_P(Q^2)$,
      obtained using the lattice data for $G_A^{u-d}(Q^2)$ and
      PPD (open circles) as well as using two two-state
      fit approaches to obtain $G^{u-d}_P(Q^2)$, namely using $M1$
      (crosses) in which we fit the temporal component of the axial
      current, i.e. $\Pi^A_0$ in Eq.~\ref{eq:ffs}, and use the same
      fit parameter for $E_1$ in the two- and three-point function,
      and $M2$ (filled circles) in which we fit the same component but
      allowing $E_1$ to be different in two- and three-point
      function.}\label{fig:ppd}
  \end{minipage}
\end{figure}

The isovector axial form factors of the nucleon are shown for the two
\Nf{2} ensembles and for \texttt{cB211.64} in Fig.~\ref{fig:GAv}. The
two \Nf{2} volumes allow for a comparison of volume effects, for which
do not observe any at the accuracy of out data. Furthermore, both
\Nf{2} ensembles with $a$=0.0937~fm are compatible with the finer
\Nf{2}{1}{1} ensemble with $a$=0.0801~fm. We note that the induced
pseudoscalar form factor shows a large discrepancy compared to the
expectation from pion pole dominance (PPD), shown with the open
symbols in Fig.~\ref{fig:ppd} for ensemble \texttt{cB211.64}. PPD
relates the induced pseudoscalar form factor with the axial form
factor, $G^{u-d}_P(Q^2) = \frac{4m_N^2}{Q^2+m_\pi^2}G^{u-d}_A(Q^2)$,
where $m_N$ and $m_\pi$ are the masses of the nucleon and pion
respectively and the $u-d$ superscript is used to denote that this
applies for the isovector case.

In Ref.~\cite{Jang:2019vkm}, it was suggested that this discrepancy may be
due to excited state contamination and that the use of the temporal
component of the axial current, typically unused due to its high
contamination from excited states, can be fitted to identify their
contribution and thus isolate, via multi-state fits, more precisely
the ground state contribution. From our analysis on the
\texttt{cB211.64} ensemble, shown in Fig.~\ref{fig:ppd}, we find that
indeed the fit yields higher values of $G^{u-d}_P(Q^2)$ but not
sufficient for agreement with the PPD. Further analysis is being
carried out to identify other sources of systematic uncertainties,
such as additional excited state contamination and cut-off effects.

\begin{figure}[h]
  \includegraphics[width=0.465\linewidth]{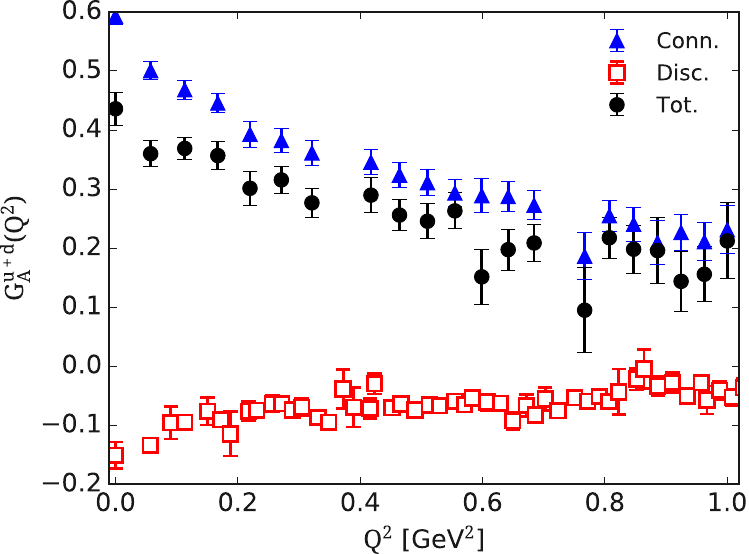}
  \hfill
  \includegraphics[width=0.465\linewidth]{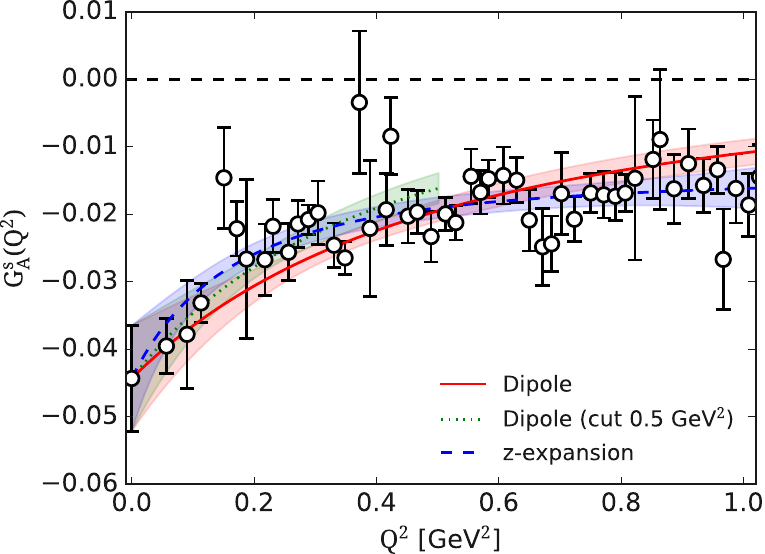}
  \caption{The isoscalar (left) and strange (right) contributions to
    the nucleon axial form factor. In the left panel we plot
    separately the connected (blue triangles) and disconnected (red
    squares) contributions as well as there sum (black circles). In
    the right panel, we also show fits to the dipole form, either
    fitting all data points (solid red line) or up to
    $Q^2=0.5$~GeV$^2$ (dotted line) as well as with the $z$-expansion
    (dashed line).}\label{fig:isos}
\end{figure}

\begin{figure}[h]
  \centering\includegraphics[width=\linewidth]{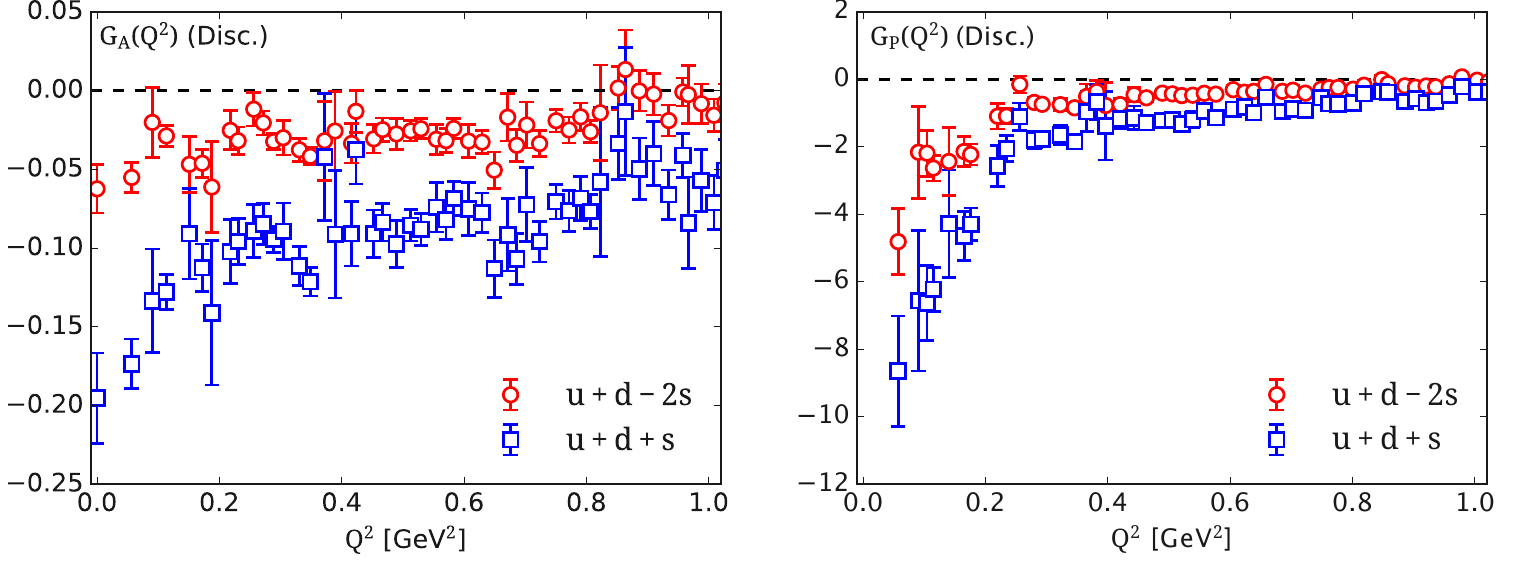}
    \caption{Disconnected contributions to the singlet (red circles)
      and octet (blue squares) combinations of the nucleon axial
      (left) and induced pseudoscalar (right) form factors for the
      \texttt{cB211.64} ensemble.}\label{fig:GAPdisc}
\end{figure}

In Fig.~\ref{fig:isos}, we show the isoscalar and strange contributions
to the nucleon axial form factor. For the isoscalar case, we plot the
connected and disconnected contributions separately in addition to
their sum, and observe that the disconnected contributions are
negative, lowering the values obtained from the connected
contributions, and significant, in contrast to the electromagnetic
case. The strange axial form factor at zero momentum transfer yields
the strange axial charge, for which we find
$g^{s}_A{=}G_A^s(Q^2{=}0){=}-0.044(8)$, and is of phenomenological
significance since it yields the contribution of the strange quark
intrinsic spin to the nucleon spin. The bands show fits to the
$z$-expansion, as well as to the dipole form,
$G_A^s(Q^2){=}G_A^s(0)/[1+\frac{Q^2}{(m_A^s)^2}]^2$.  The radius is
obtained from the dipole mass via $\langle r^2_A \rangle^s =
12/(m^s_A)^2$. With the isoscalar, isovector, and strange quark
contributions to the axial form factors at hand, we can construct the
phenomenologically interesting SU(3) singlet $u{+}d{+}s$ and octet
$u{+}d{-}2s$ combinations. The disconnected contributions to these
combinations are shown in Fig.~\ref{fig:GAPdisc}. We find that both
combinations, for both axial and induced pseudoscalar form factors,
are non-zero and negative. We note that in the SU(3) flavor symmetric
limit, the octet combination should be zero, and therefore any
deviation signals SU(3) flavor symmetry breaking.

\begin{figure}[h]
  \includegraphics[width=0.475\linewidth]{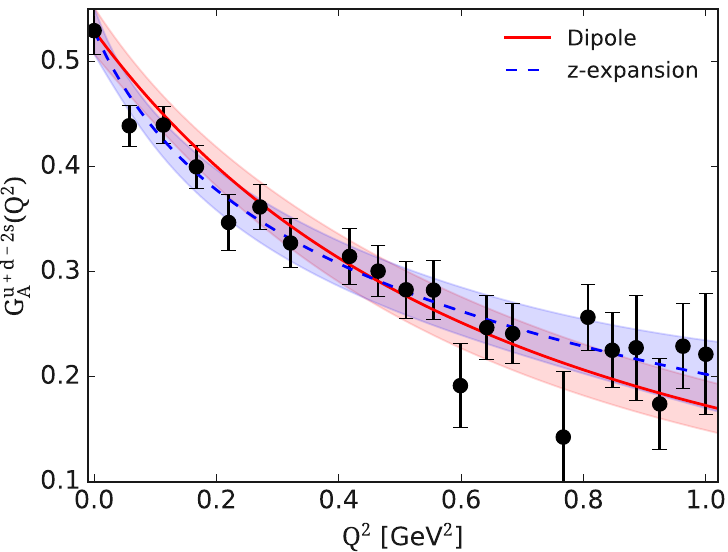}\hfill
  \includegraphics[width=0.475\linewidth]{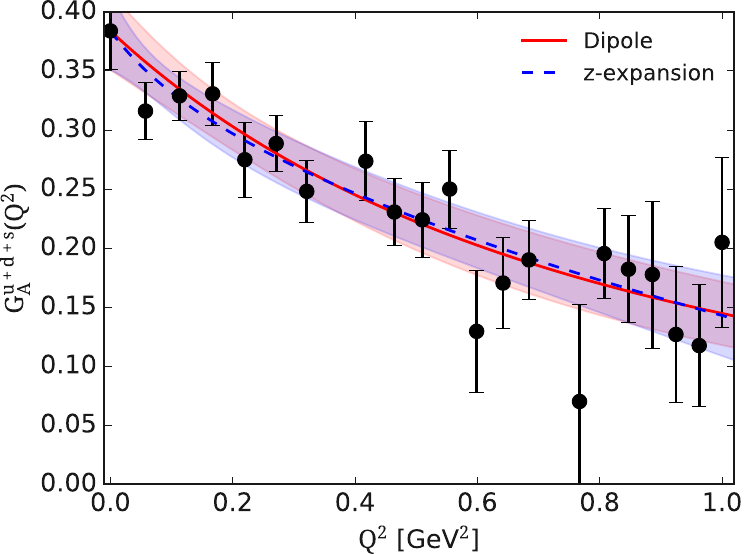}
  \caption{The singlet (left) and octet (right) combinations of the
    nucleon axial form factor. We show fits to the dipole (solid red
    curve) and $z$-expansion (dashed blue
    curve).}\label{fig:GAsinglet-octet}
\end{figure}

The singlet and octet combinations of the nucleon axial form factor
are shown in Fig.~\ref{fig:GAsinglet-octet}. The inclusion of
disconnected contributions and physical point simulations is carried
out for the first time for these quantities. We fit to both dipole and
$z$-expansion forms to obtain the radii and moments. In
Table~\ref{tab:radii} we tabulate results for the radii. Among
$z$-expansion and dipole fits and the different momentum transfer
cuts, we obtain consistent results within the statistical
uncertainties. More details on the fits can be found in
Ref.~\cite{Alexandrou:2021wzv}.

\begin{table}
  \caption{Fit results for the axial singlet, octet, and strange radii
    (denoted $si$, $oc$, and $s$). We show results with a dipole or
    $z$-expansion form and using either data up to $Q^2\simeq
    1$~GeV$^2$ or $Q^2\simeq 0.5$~GeV$^2$.}\label{tab:radii}
  \begin{tabular}{ccr@{.}lcr@{.}lcr@{.}lc}
    \hline\hline
    Fit Type    & $Q^2_{\rm max}$ [GeV$^2$]                         & \multicolumn{2}{c}{$\sqrt{\langle r_A^2 \rangle^{si}}$ [fm]} & $\frac{\chi^2}{\textrm{d.o.f}}$ & \multicolumn{2}{c}{$\sqrt{\langle r_A^2 \rangle^{oc}}$ [fm]} & $\frac{\chi^2}{\textrm{d.o.f}}$ & \multicolumn{2}{c}{$\sqrt{\langle r_A^2 \rangle^s}$} [fm] & $\frac{\chi^2}{\textrm{d.o.f}}$ \\
    \hline
    \multirow{2}{*}{Dipole}      & $\simeq$ 0.5                             & 0                                                           & 623(59)                                          & 1.07                                                       & 0                                                & 545(104)                                             & 0.68 & 0 & 782(145) & 1.33                       \\
                                 & $\simeq$                               1 & 0                                                           & 592(52)                                          & 1.04                                                       & 0                                                & 542(81)                                              & 0.65 & 0 & 689(114) & 1.48                       \\
    \multirow{2}{*}{z-expansion} & $\simeq$ 0.5                             & 0                                                           & 780(108)                                         & 0.45                                                       & 0                                                & 673(221)                                             & 0.50 & 0 & 973(248) & 0.99                       \\
                                 & $\simeq$                               1 & 0                                                           & 761(113)                                         & 0.57                                                       & 0                                                & 650(221)                                             & 0.59 & 0 & 984(239) & 0.81                       \\
    \hline\hline
  \end{tabular}
  \vspace{-2ex}
\end{table}

\section{Summary and outlook}

The electromagnetic and axial form factors of the nucleon are
calculated using three lattice ensembles with physical pion mass at
multiple sink-source separations. The two \Nf{2} ensembles are
simulated at two volumes and allow for studying volume effects. Our
comparisons show that for the isovector case no such effects are
observed within our statistics. Disconnected diagrams are calculated
for an \Nf{2} ensemble with $a$=0.0937~fm and an \Nf{2}{1}{1} ensemble
with $a$=0.0801~fm, which allow the determination of the proton and
neutron form factors, and the strange form factors. By fitting the
$Q^2$-dependence of these form factors we can extract phenomenological
quantities of interest, such as the strange axial charge and
associated radii of the nucleon.  Analysis is under way using physical
point \Nf{2}{1}{1} ensembles at finer lattice spacings to evaluate
cut-off effects. Preliminary results are shown here for the strange
electromagnetic form factors using an ensemble with
$a{\simeq}0.07$~fm. The additional ensembles will allow taking the
continuum and infinite volume limits, while additional analyses are
planned to further identify systematic errors caused by remaining
excited state contamination.\vspace{-2ex}

\acknowledgments\vspace{-2ex} We thank all members of the ETM
collaboration for a most conducive cooperation. G.K. acknowledges
support from project NextQCD, co-funded by the European Regional
Development Fund and the Republic of Cyprus through the Research and
Innovation Foundation (RIF) (\texttt{EXCELLENCE/0918/0129}). S.B. and
J.F. are supported by the H2020 project PRACE 6-IP (GA No. 82376) and
the EuroCC (GA No. 951740). K.H. is supported by RIF under contract
no.  \texttt{POST-DOC/0718/0100}.  A.V. is supported by the U.S. NSF
under Grants No. PHY17-19626 and PHY20-13064.  Partial support is
provided by the Marie Sklodowska Curie joint doctorate program
STIMULATE (GA No. 765048).  We acknowledge the Gauss Centre for
Supercomputing e.V. (\url{www.gauss-centre.eu}) for project
\texttt{pr74yo} by providing computing time on SuperMUC at LRZ
(\url{www.lrz.de}) and JUWELS-booster at the JSC
(\url{www.fz-juelich.de}). Results were obtained using Piz Daint at
CSCS, via the projects with ids \texttt{s702} and \texttt{s954}. This
work used resources from NIC on JUWELS at the JSC, under projects with
ids \texttt{ECY00} and \texttt{HCH02}. We acknowledge PRACE for
awarding us access to Marconi100 at CINECA, Piz Daint at CSCS, and
HAWK at HLRS, where part of our work is carried out within the project
with Ids \texttt{Pra20\_5171}, \texttt{pr79}, and \texttt{Acid 4886}.

\providecommand{\href}[2]{#2}\begingroup\raggedright\endgroup

\end{document}